# Space-time crystals from particle-like topological solitons

Hanqing Zhao[1,2] and Ivan I. Smalyukh[1,2,3,4*]

[1]*Department of Physics, University of Colorado, Boulder, CO 80309, USA*

[2]*International Institute for Sustainability with Knotted Chiral Meta Matter (WPI-SKCM²), Hiroshima University, Higashi Hiroshima, Hiroshima 739-8526, Japan*

[3]*Materials Science and Engineering Program, University of Colorado, Boulder, CO 80309, USA*

[4]*Renewable and Sustainable Energy Institute, National Renewable Energy Laboratory and University of Colorado, Boulder, CO 80309, USA*

*\* Correspondence to:  ivan.smalyukh@colorado.edu*

**Abstract** Time crystals are unexpected states of matter that spontaneously break time translation symmetry either in a discrete or continuous manner. However, spatially-mesoscale space-time crystals that break both the space and time symmetries have not been reported. Here we report a continuous space-time crystal in a nematic liquid crystal driven by ambient-power, constant-intensity unstructured light. Our numerically constructed 4-dimensional configurations exhibit good agreement with these experimental findings. While meeting the established criteria to identify time-crystalline order, both experiments and computer simulations reveal a space-time crystallization phase formed by particle-like topological solitons. The robustness against temporal perturbations and spatiotemporal dislocations shows the stability and rigidity of the studied space-time crystals, which relates to their locally topological nature and many-body interactions between emergent spontaneously-twisted, particle-like solitonic building blocks. Their potential technological utility includes optical devices, photonic space-time crystal generators, telecommunications, and anti-counterfeiting designs, among others.

Highly ubiquitous crystals and liquid crystals (LCs) have been driving major scientific discoveries and technological innovation for centuries[1,2]. Time crystals were proposed by Wilczek only a decade ago[3–5], with the initial concept shown later to be unachievable[6,7]. However, new concepts of time crystals recently captured the fascination of numerous researchers who are now contributing to this explosively growing field[8–13]. Differing from Wilczek's designs of time crystals based on closed quantum and classical many-body systems[3,4], time crystals with external Floquet drives that break time translation symmetry discretely, dubbed "discrete (Floquet) time crystals"[14–17], have been demonstrated. The quantum discrete time crystals have been observed in systems of nuclear spins, trapped ions, cold atoms, superconducting qubits and so on[8–13,18–28]. Recently, a continuous time crystal (which does not rely on periodic Floquet driving) has been observed in a quantum system of atom cavity[29], also breaking the time translation symmetry and opening a new avenue for the studies of time crystals.

To differentiate the genuine time-crystalline order from a large variety of time-periodic patterns, stringent requirements have been introduced[29]. First, these crystals should emerge from spontaneous time symmetry breaking, where the relative time phases are randomly distributed between 0 and $2\pi$, assuring that the time crystals are independent of the external source. Second, time crystals should exhibit robustness against temporal perturbations by an external source, demonstrating a rigidity analogous to that of space crystals. Criteria-satisfying continuous time crystals have been reported for photonic metamaterials[30], optically pumped atomic system[31], Rydberg gases[32] and electron-nuclear spin systems[33]. However, continuous space-time crystals (CSTCs), which spontaneously break the time translation symmetry as well as the space translation symmetry without assistance of periodic driving, have not been convincingly demonstrated yet in either quantum or classical systems[29–34]. Moreover, since most of the time crystals known so far exist in the quantum world, a time crystal has not been directly seen by microscopic observations or even with bare human eyes, at the same time being shown to meet the above identification criteria. While time-varying patterns are ubiquitous and can often be observed by naked eyes, they are typically not arising from many-body interactions of quasi-atom building blocks, and whether some of them could happen to meet the stringent requirements (described above) of being identified as time crystals remains an open question[9,11,12,31].

Here we report the observation of classical CSTCs in nematic LC systems with emergent topological soliton building blocks exhibiting interactions mediated by orientational elasticity. By shining ambient or microscope illumination light on specially designed samples with LC confined between glass plates coated by photo-responsive dye, we find emergent spatiotemporal patterns in which spatial and temporal symmetries are broken, without the assistance of periodic external driving. Furthermore, such emergent behaviour can be directly observed in an optical microscope and, when designed appropriately, even by bare eyes. The continuous space-time crystallization phase is quite stable at room temperature and can persist locally for hours of our observation. By building a model based on balancing optical, surface and bulk viscoelastic torques associated with nematic fluid's orientational viscoelasticity, we explain the emergent formation of topological solitons in this system. We then construct the periodically-varying configurations with elastically interacting spatially-localized particle-like topological solitonic structures. We find that the intrinsic temporal periodic nature persists with changes of temperature and external driving light intensity, yielding good agreements with the experimental observations. We also verify experimentally and numerically that the time translation symmetry is broken spontaneously and the continuous space-time crystallization phase is robust against spatiotemporal dislocations and temporal perturbations, meeting all requirements to be identified as time-crystalline. Finally, we discuss implications of our work for fundamental science and technological utility in optical devices, photonic space-time crystal generators, telecommunications, anti-counterfeiting designs, and cryptography, among many other.

**Emergence of the time-crystalline order**

The samples are prepared by sandwiching the nematic LC with photo-responsive confining substrates, inner surfaces of which are coated with dye. The photo-responsive azobenzene dye molecules (Methods) at the surfaces and LC molecules in the bulk have anisotropic shapes[2,35], with their average orientations characterized by nonpolar directors $\mathbf{n_s}$ and $\mathbf{n}$ with head-tail symmetry ($\mathbf{n_s} \equiv -\mathbf{n_s}$ and $\mathbf{n} \equiv -\mathbf{n}$), respectively. When normally incident linearly polarized blue light passes through the samples (Fig. 1a), $\mathbf{n_s}$ at the top surface tends to orient perpendicularly to the polarization direction of the linearly polarized light, and the LC molecules near it orient parallel to the azobenzene molecules, thus following the dye-defined surface boundary conditions (Fig. 1b-d). For a monodomain LC and polarization direction of the linearly polarized light orthogonal to directors $\mathbf{n_s}$ at both top and bottom surfaces and $\mathbf{n}$ in the bulk, the light could traverse the sample as an ordinary mode without polarization change. However, the initial surface boundary conditions are tangentially degenerate. Upon the illumination of linearly polarized light, the dye molecules at the top surface become uniformly oriented along the direction perpendicular to the polarization of incident light, whereas orientations of the dye molecules at the bottom substrate depend on their initial random polydomain orientations. Orientational changes of the LC director develop across the sample thickness, leading to transformation of the polarization state of the traversing light to generally elliptical polarization, as well as rotation of the long axis of polarization ellipse. Subsequently, as the bottom substrate is exposed to the elliptically polarized light, this leads to the reorientation of neighbouring LC molecules and $n_s$ at the bottom surface, which tend to orient perpendicular to the major axis of the polarization ellipse of elliptically polarized light (Fig. 1d). Once the dye molecules and $n_s$ rotate, they also drive reorientations of the neighbouring LC molecules and **n**, further changing the polarization ellipse of traversing polarized light. Such feedback mechanism spontaneously creates periodic spatiotemporal array of topological solitonic quasiparticles (Fig. 1e and Supplementary Videos 1,2), forming a robust continuous space-time crystallization phase (Fig. 1f and Extended Data Fig. 1), which will be described in detail below.

With the help of the first-order full-wave retardation plate (Fig. 1a), the different orientations of the LC director in the bulk are inferred from polarized interference colours seen in the captured polarizing optical micrographs. For a thin cell (thickness $d$<2μm), the light after the analyser becomes blueish or yellowish (Fig. 1e) when the director orientations are parallel or perpendicular to the slow axis of the retardation plate, respectively. The spatial area of an CSTC can be made larger than 1 mm$^2$, which can be observed directly by eyes and can be characterized in detail with an optical microscope. By selecting a fixed stripe area (dashed rectangle in Fig. 1e) and tracking it in time, a space-time image of the solitonic arrays is composed, where the size of a representative time crystal region shown in Fig. 1f is 400μm·120s. We analyze the normalized digital intensity signals ($\Phi$) of pixels in Fig. 1f and perform the Fast Fourier Transform analysis (Fig. 1g,h), showing a sharp peak at around 0.217 Hz (4.61s), which corresponds to the temporal periodicity of the CSTC. By defining time correlation function $G(t) = \langle\Phi(t)\Phi(0)\rangle-\langle\Phi(t)\rangle\langle\Phi(0)\rangle$ from the normalized digital intensity signals within the pattern (Methods), we find that $G(t)$ can be fitted by a power-law decay (~$t^{-0.09}$) despite of fluctuations, exhibiting a quasi-long-range order in time (Extended Data Fig. 2), similar to that known for systems with one-dimensional spatial positional order like smectic LCs that follow the Peierls and Landau's predictions[36].

**Numerical modeling of emergent particle-like solitonic structures and crystals**

To gain insights into the emergent formation of particle-like topological solitons that then exhibit time crystallization, we use a model based on the balance of optical, surface anchoring and bulk viscous and elastic torques, with the latter derived from the Frank-Oseen free energy[2,37]:

$$F_{\text{bulk}} = \int d^3\boldsymbol{r} \left\{ \frac{K_{11}}{2} (\nabla \cdot \boldsymbol{n})^2 + \frac{K_{22}}{2} [\boldsymbol{n} \cdot (\nabla \times \boldsymbol{n})]^2 + \frac{K_{33}}{2} [\boldsymbol{n} \times (\nabla \times \boldsymbol{n})]^2 \right\}, \quad (1)$$

where the Frank elastic constants $K_{11}$, $K_{22}$ and $K_{33}$ depend on temperature (Methods, Supplementary Fig. 1 and Supplementary Tables 1,2) and determine energetic costs of, respectively, splay, twist and bend spatial deformations of the LC director field, **n(r)**. Furthermore, the free energy term associated with coupling **n(r)** and **n$_s$(r)** at surfaces reads

$$F_{\text{surface}} = -\int d^2\boldsymbol{r} \, \frac{\eta W}{2} (\boldsymbol{n_s} \cdot \boldsymbol{n})^2, \quad (2)$$

where $\eta$ is the dimensionless light coupling efficiency parameter ranging from 0 to 1 and $W$ is the anchoring coefficient. We assume that **n$_s$** orients perpendicular to the direction of major axis of the elliptically polarized light, calculated using the Jones-matrix method (Methods). While our field theory model is intended to describe spatial structures of the director field, solitonic topological particles emerge spontaneously within it, much like skyrmions, domain walls and hopfions found in diverse fields ranging from magnets to LCs[40]. In turn, these particle-like solitonic objects then exhibit many-body interactions and the time-crystalline order.

Consider a uniform state of director field with a small perturbation on the left side (Extended Data Fig. 3a), when light (450nm) normally incident on the sample from the top has linear polarization along $y$ axis. The resulting director dynamics stems from the torque balance $\frac{\delta F}{\delta \mathbf{n}_i} = -\gamma \frac{\partial \mathbf{n}_i}{\partial t}$ yielding the temporal evolution **n**$_i$(t), where $\frac{\delta F}{\delta \mathbf{n}_i}$ is the variational derivative of $F$, subscript *i* denotes spatial coordinates, $\gamma$ is the rotational viscosity, Jones vectors are updated further in each next iteration and $F=F_{bulk}+F_{surface}$. A spatially periodic configuration of the solitonic array emerges spontaneously (Fig. 2 and Extended Data Fig. 3). By properly arranging the initial state for periodic boundary conditions (Methods), we find that the simulated structure shown in Fig. 2a exhibits properties of a continuous space-time crystal phase (Supplementary Videos 3,4).

In the dynamic steady-state configuration, the director field **n(r)** at the top surface stays perpendicular to the incident light's polarization direction and smoothly deforms at the bottom (Fig. 2a), displaying alternating pairs of the nematic Néel domain wall solitons of opposite elementary topological charge[1,38,39] (Fig. 2a,b). These solitons are labelled as the elements of the first homotopy group $\pi_1(\mathbb{S}^1/\mathbb{Z}_2)\equiv\pi_1(\mathbb{S}^1)=\mathbb{Z}$ and can be treated as topological quasiparticles of opposite signs ±1, low-dimensional analogues of Skyrme solitons used to model particles with different baryon numbers in subatomic physics[38,40]. Treating the time coordinate similarly to the space coordinates, ±1 elementary quasiparticle-like topological solitons can be also identified in time. By calculating the elastic free energy of the LC director field deformations above the solitonic walls in the dye orientations and the adjacent LC director field (Fig. 2a-c), we find that elasticity-mediated interactions between the solitonic quasiparticles can be equivalently represented with the help of topological elastic bonds connecting the neighbouring interacting quasiparticles (Fig. 2d), forming a many-body interaction system. Because the entire system is topologically neutral (no net topological charge), individual quasiparticles cannot be smoothly eliminated but rather only can vanish through annihilating pairs of them, which would require a large energetic barrier. By analyzing the relative displacements distribution of neighbouring +1 Néel domain wall solitons associated with different spatial lattices at different times in the space-time crystal, we find that the relative displacements follow a Gaussian distribution (Fig. 2e), corresponding to an effective harmonic potential energy landscape (Fig. 2f) under small thermally driven displacements. Such an energy landscape describing the quasiparticle interactions resembles that found in a colloidal particle crystal systems in liquid crystal media [41], where the energy differences are several $k_BT_{em}$ corresponding to ~ 10% displacements ($k_B$ is the Boltzmann constant and $T_{em}$ is the temperature). It indicates that, albeit all emergent from and describable by the field theory of the LC director field, many-body interactions between the quasiparticles maintain the order of space-time crystals.

The close agreements between computer-simulated and experimental polarizing optical micrographs (Fig. 1e, Fig. 2g and Extended Data Fig. 4a,b) and three-photon excitation fluorescence polarizing microscopy images (Extended Data Fig. 4c,d) validate our analysis of CSTC configurations. Similarly, the theory-experiment agreements of space-time polarizing optical micrographs and videos support the reconstructed configuration of the CSTC (Fig. 1f, Fig. 2h and Supplementary Videos 1,4).

**Diversity and control of the space-time periodic structures**

In addition to the emergence of time-space-periodic domain wall arrays, by superimposing the light patterns from two CSTCs, we find that the overlapping spatial solitonic arrays can allow us to form orthorhombic and monoclinic lattices. The corresponding combined space-time patterns exhibit the same space-time lattices (Extended Data Fig. 5). Due to the non-equivalence between spatial and temporal coordinates, we only find the 1+1 dimensional (1+1D) space-time orthorhombic and monoclinic lattices (intrinsically different from the 2D wallpaper space groups), consistently with previous theoretical proposals[42]. This result reveals the possibility of achieving a richer space-time group structure in higher dimensions, as obtained here by incorporating the time coordinate into the spatial symmetry group[42].

External stimuli allow for controlling the studied emergent spatiotemporal order. The temporal periodicity of CSTC increases as the driving light intensity decreases (Methods, Fig. 2i, Extended Data Fig. 6 and Supplementary Video 5). Below a certain threshold (~1 mW cm$^{-2}$), the continuous space-time crystallization phase transforms to a time-symmetry-unbroken (TSU) phase, where the LC director field becomes static. Computer simulations qualitatively reproduce this behaviour as the temporal periodicity decreases with increasing $\eta$ (Fig. 2j and Supplementary Fig. 2), and the structure eventually relaxes to the spatially uniform planar state (Extended Data Fig. 7) at efficiencies below a certain threshold ($\eta$~0.3). The temporal periodicity also decreases with increasing temperature, until the system reaches a disorder phase (Fig. 2i,j). Noteworthily, the Frank elastic constants and birefringence of the LCs decrease by more than 30% within such a temperature change[37], indicating a good robustness of the CSTCs that can withstand changes of these material parameters. Experimental observations with changing temperature and driving light intensity in the same sample are consistent with computer simulations that yield spatial periodicity relatively insensitive to temperature and light coupling efficiency. For different LC cells with high birefringence and varying thickness, the temporal periodicities can be controlled to range from tens of seconds to milliseconds (Supplementary Fig. 3). Second-scale time crystals could be employed for interfacing with biological and organic holographic materials, which, in turn, can be potentially used for optical signal amplification, data storage, and phase conjugation[43].

**Testing against stringent criteria for time-crystalline identification**

Meeting a key criterium of time crystallinity[29,30], our space-time crystals exhibit good robustness against spatial and temporal perturbations (Fig. 3). The CSTCs are found to recover defect-free order within tens of temporal periods after an emergence of space-time dislocation (Fig. 3a). Interestingly, by reconstructing the space-time coordinates, we find that the profiles after the dislocation emergence can be fitted by the nonlinear theory of an edge dislocation in smectic crystal systems (Fig. 3b), showing similarity of spatial and temporal quasi-particle displacements resembling that of molecules in smectics in higher dimmensions[44]. The space-time crystals are also stable under temporal perturbations, to confirm the robustness against temporal perturbations (Fig. 3c-i), we use blue light with temporally randomized intensity (randomly distributed within the range $[\alpha,1]W_{driving}$, where $\alpha$ is the parameter controlling the strength of perturbation and $W_{driving}$ is the maximum driving light intensity) to illuminate the sample and red light in a separate channel for imaging (Supplementary Fig. 4), the continuous space-time crystallization order persists well under small temporal perturbations of incident light intensity ($\alpha = 0.8$ and $W_{driving} = 1.5$ mWcm$^{-2}$). The effect of temporal perturbations can be quantified by measuring the crystalline fraction parameter, $\varXi$[19,29]. Both experiments and computer simulations show that the crystalline fraction does not change much when the perturbation within a certain range (Fig. 3f-i), which corresponds to the space-time crystal phase. As $\alpha$ decreases, the perturbation becomes larger and $\varXi$ suddenly drops, because the system experiences a phase transition and becomes disordered (Fig. 3g,i).

As anticipated for space-time crystals, the continuous time translation symmetry of CSTCs is spontaneously broken independently of the driving light source. To reveal this, we repeatedly measure the time phase each time after blocking and shining the blue light to a CSTC area (Fig. 4a,b), finding the relative time phases randomly distributed between 0 and $2\pi$ (Fig. 4c), as expected[29,30,32,33]. By blocking and unblocking the external light source and tracking the director field (Fig. 4d,e) we also reproduce this general behaviour in computer simulations that closely agree with experiments.

**Comparison to other classical time-periodic effects**

We have described the observation of classical CSTCs in nematic LCs, which emerge at room temperature and can be designed to be seen by optical microscopes or even directly by human eyes. These spatiotemporal states usually emerge within an area (Extended Data Figs. 3,8 and Supplementary Video 6) where the one-dimensional spatial symmetry and temporal symmetry are spontaneously broken in a continuous manner, forming 1+1D CSTCs. This emergent behaviour is different from the case of active (moving) crystals, where a pre-set crystal structure can spatially translate while periodically entering similar states of the spatially periodic lattice, driven by external electric signals[45]. Breaking of temporal symmetry of these active crystals refers to the breaking of spatial translation symmetry, and the active lattice time periodicity depends on the collective moving speed[45] of spatial crystallites, differently from our time crystals.

CSTCs persist locally for hours while the temporal periodicity changes only slightly (by ~10%) after ~1hour of observations, where properties of LCs largely define their behavior. The emergent bonded quasiparticles reveal robustness against temporal and spatial perturbations, demonstrating intrinsic time crystallinity features and meeting identification criteria that so far have not been probed for other classical temporally periodic systems like chemical oscillators and dissipative structures (potential candidates for time crystallinity of physical behavior, which remain to be tested against the identification criteria), whereas our CSTC is a classical topological solitonic system that meets all time crystal's stringent requirements introduced so far[29–33,46].

While we focused on 1+1D CSTCs, by introducing higher dimensional topological solitons or defects and their arrays [39,47,48], 2+1D and 3+1D and other CSTCs can be potentially observed. Furthermore, the demonstrated pre-designed realization of continuous time crystals in classical soft matter systems may stimulate efforts towards realizing discrete time and space-time crystals, where facile responses of LCs to external stimuli can be harnessed as well. Overall, unambiguous demonstrations of time-crystalline physical behavior in highly technological LC materials has the potential of harnessing the exciting developments and unexpected observations in time crystalline systems to enable their technological utility, as we illustrate with examples below.

**Technological potential**

CSTCs in highly technological LC materials promise applications in optical devices, photonic space-time crystal generators, and telecommunications. Since the spatial structures of CSTCs consist of the Néel domain wall solitons, when polarized light passes through the nematic LC slab with such topological solitons at one surface and periodic director deformations extending throughout the nematic bulk, the traversing polarized light accumulates phase retardation according to the director's spatially and temporally varying orientations (Fig. 5). This property may enable the fabrication of dynamic time-crystalline Pancharatnam-Berry phase (geometric phase) gratings and lenses (also known as cycloidal diffractive waveplates)[49,50]. We note that the phase accumulations vary not only with spatial coordinates but also with temporal coordinates, where the outgoing light's polarization state depends on the wavelength and polarization of the input light (Fig. 5b-d). We therefore foresee that such LC-based space-time crystals can be developed further for generating photonic space-time crystals, assuming that the spatial periodicity can be reduced to approach the visible light's wavelengths[51] and the temporal frequencies can be much higher[11]. In addition, one can combine CSTC-modulated light with the driving light, allowing CSTCs to change the polarization states of the accompanying modulated light without affecting CSTC's 4D structures. For example, as the typical wavelength range used in fiber optical telecommunications (>850nm) differs from that of the driving light (~450nm) to which azobenzene dye is sensitive, optical signal modulation and information encoding at wavelengths outside the spectral range of the azobenzene dye's sensitivity could be implemented based on the CSTC's robust temporal and spatial order (Fig. 5c,d).

The ambient-intensity light-driven and non-contact properties of CSTCs make them suitable for anti-counterfeiting[52] with the protection at multiple implementation levels in which the temporal and spatial periodic structures can spontaneously emerge (Fig. 6a-c). This "time watermark" can be fabricated at low cost, since a 1cm×1cm×2μm sample requires only ~$2\times10^{-4}$ g of LC and the surface monolayer with <$10^{-14}$g of the azobenzene dye that can be sandwiched between glass or other surfaces[35]. Due to the spontaneous temporal symmetry breaking, the time crystals can be exploited in pseudo-random number generators[30]. By combining multiple CSTCs, the synthetic systems generate unique, fingerprint-like states corresponding to the spatiotemporal topological soliton arrays (Fig. 6d,e) each time they emerge, maintaining order for a remarkably long time (Extended Data Fig. 2). Additionally, the phases of CSTCs can be tuned by smoothly switching the driving light intensity (Fig. 6f), allowing for the creation of a 2+1D barcode via superimposing multiple CSTCs (Fig. 6g and Supplementary Video 7). As the 2D barcode can store over 100 times more bits than a 1D barcode, the capacity of storing information with proper data coding in higher-dimensional barcodes like 2+1D is effectively unbounded due to the extra temporal coordinate (>100,000 bits per second)[53]. The intrinsic robustness of the space-time order, supported by the time-crystallinity of topological solitons, could further enhance error correction capabilities of 2+1D barcodes. For the third level of the anti-counterfeiting uses, the temporal periodicity of CSTCs can be utilized as keys in cryptographic systems. For example, with two CSTCs having temporal periodicities $T_1$ and $T_2$ (assuming $T_1<T_2$), an identical spatial pattern only recurs after a time interval of $(T_1 \times T_2)/(T_2-T_1)$ (Fig. 6h and Supplementary Video 5), which could be used to check authenticity. CSTCs with different temporal periodicities can be introduced by incorporating pre-programmed light intensity filters. The entire system may display disorder-like spatiotemporal behavior, however within each CSTC, the time-crystalline order with a specific temporal periodicity can be maintained for a long time. Overall, the entire fingerprint-like CSTC states can be precisely predicted if the information about the pre-programmed light intensity filters (the system's keys) is known (Fig. 6i) and can be utilized for anti-counterfeiting purposes.

**Conclusions**

Our observations of CSTCs may indicate abundance of time crystallization in soft matter systems like LCs, promoting studies of space-time groups with their topological properties[42]. While meeting the current "stringent criteria" for identifying the time crystalline order, our study of CSTCs points to the need of examining other classical time-periodic effects in various physical and chemical systems to test if they can be identified as time-crystalline, where new "identity" criteria may be needed to differentiate them. The diverse forms of true or pseudo time crystallinity will drive developments of materials with fundamentally interesting and technologically useful emergent nonlinear phenomena, like the one described here. Indeed, our findings suggest that the spontaneous symmetry breaking both in time and space can be a widespread occurrence in numerous open systems, whether in a quantum or in a classical context. On the other hand, the fact that time crystal states can occur in LCs with myriads of technological uses under conditions compatible with many technological needs may suggest new breeds of electro-optic and photonic devices[54,55]. Potential technological applications in optical devices, photonic space-time crystal generators, telecommunications and anti-counterfeiting designs may signal the beginning of an exciting frontier for the time and space-time crystals, where fundamental research advances could drive technological utility.

**Acknowledgements:** We thank Taewoo Lee for technical assistances, Lech Longa, Boris Malomed and Lingzhen Guo for discussions, as well as Qingkun Liu and Ye Yuan for synthesizing the azobenzene dye molecules. I.I.S. acknowledges hospitality of the International Institute for Sustainability with Knotted Chiral Meta Matter (SKCM$^2$) in Japan while he was partly working on this article while on a sabbatical. This research was supported by the US Department of Energy, Office of Basic Energy Sciences, Division of Materials Sciences and Engineering, under contract DE-SC0019293 with the University of Colorado at Boulder.

**Author Contributions:** H.Z. performed experiments and numerical modelling under the supervision of I.I.S. I.I.S. initiated and directed the research. H.Z. and I.I.S. wrote the manuscript.

**Competing Interests:**

The authors declare the following competing interests: H.Z., and I.I.S. filed a patent application related to continuous space-time crystals, as submitted by the University of Colorado concurrently with this paper.

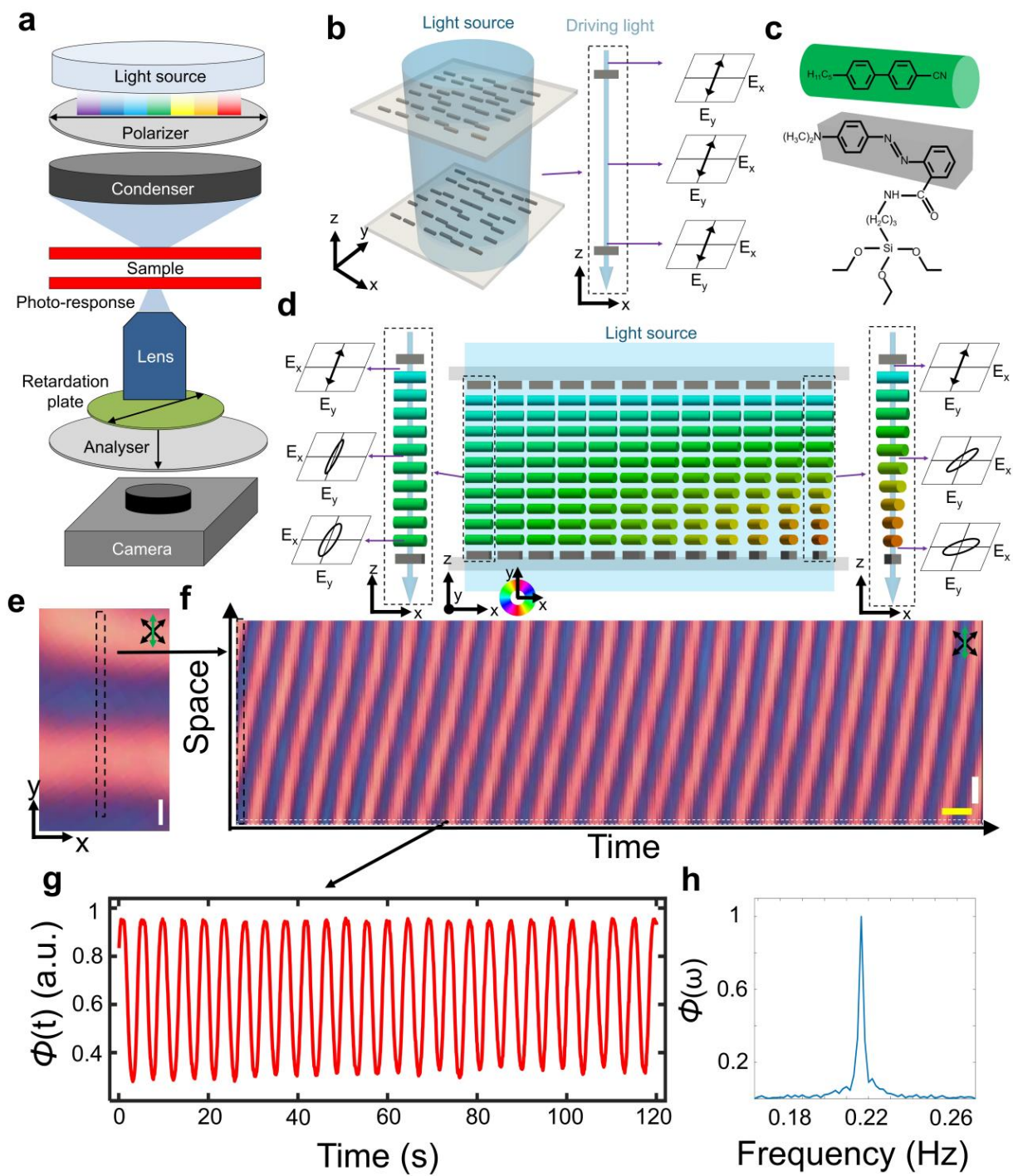

**Fig. 1 | Observation of a topological solitonic CSTC in a nematic LC system. a**, Schematic of a used optical microscope: the linearly polarized light passes through the LC glass cell, where the

inner surfaces of substrates are coated with photo-responsive dye. The light signals and spatial patterns are recorded by a camera after passing through the first-order retardation plate and an analyser. **b**, Schematic of an empty cell with normally incident linearly polarized blue light passing through. The dye molecules (gray cuboids) have long axes perpendicular to the polarization of the linearly polarized light. The electric field ($E_x$,$E_y$) of driving light at different depths is shown on the right. **c**, Chemical structures of the azobenzene dye and LC molecules. The LC molecule is represented as a green cylinder and the azobenzene group is shown as a gray cuboid. **d**, Schematic of a LC-filled sample, where LC molecules are coloured according to azimuthal angles of the director orientation, as defined by the bottom coloured scheme. The electric field ($E_x$,$E_y$) of driving light at different depths of the marked areas is plotted on the right and left, respectively. **e**, Experimental polarizing optical micrograph of the CSTC obtained with a first-order full-wave retardation plate. The plate's slow axis is labelled by the green double arrow and crossed polarizers are labelled by black double arrows. **f**, Space-time image of the CSTC shown in (e), where the crystal size is 400μm·120s. The selected area is marked in (e); time interval is 0.3s. White scale bars indicate 50μm in (e,f) and yellow scale bar indicates 5s in (f). **g**, Normalized red light signals extracted from space-time plot, where the selected area is marked in (f). a. u., arbitrary units. **h**, Normalized Fast Fourier Transform spectrum of the light signals from (g), $\Phi(t) \rightarrow \Phi(\omega)$; the central peak is at 0.217 Hz.

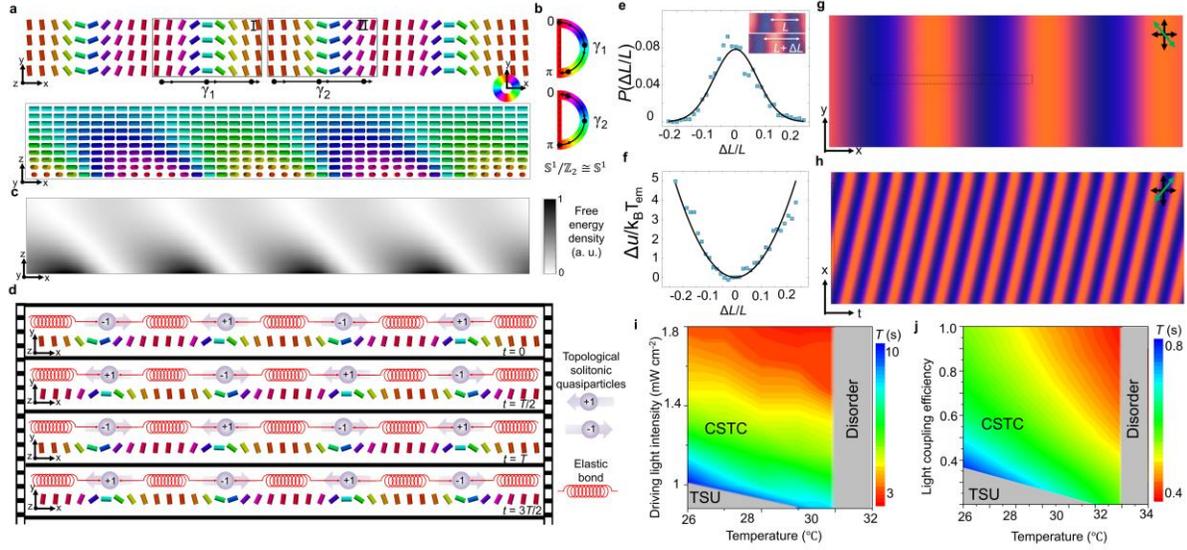

**Fig. 2 | Spatial configuration, temporal periodicity and many-body interactions of CSTCs.**
**a**, Director field **n(r)** of a CSTC in *x-y* (top) and *x-z* (bottom) cross-sections. **n(r)** is translationally invariant along the *y* axis, as illustrated *x-y* cross-section at the bottom surface. Light propagation directions are from *z* to *-z*; the cylinders are coloured according to the circle scheme on the right. **b**, Along the spatial paths $\gamma_1$ and $\gamma_2$ marked in (a), the director **n** rotates by $+\pi$ and $-\pi$, respectively. Since $\mathbb{S}^1/\mathbb{Z}_2 \cong \mathbb{S}^1$, the mappings of director orientations from these paths cover the order parameter space once in opposite directions, indicating the +1 and -1 Néel domain wall solitons at region I and II in (a). **c**, Elastic free energy density in the *x-z* cross-section marked in (a). The free energy density is calculated using Eq. (1) and relative to that of a uniform nematic background. It is visualized using grayscale scheme (right-side inset); a.u., arbitrary units. **d**, Schematic of many-body interactions among the topological solitonic quasiparticles at different times, where *T* is the temporal periodicity of the CSTC. **e,f**, Probability distributions of displacements of neighbouring domain wall solitons (e) and corresponding potential landscape (f); inset shows measurements of displacements, where *L* is the spatial periodicity. **g,h**, Simulated polarizing optical micrograph snapshot (g) and space-time image (h) of the CSTC. The selected area (h) is marked in (g). **i**,

Experimental measured temporal periodicity as a function of the temperature and driving light intensity. **j**, Simulated temporal periodicity versus temperature and light coupling efficiency; temporal periodicities $T$ are coloured according to the schemes in right-side insets whereas regions of disorder and TSU phases are shown in grey.

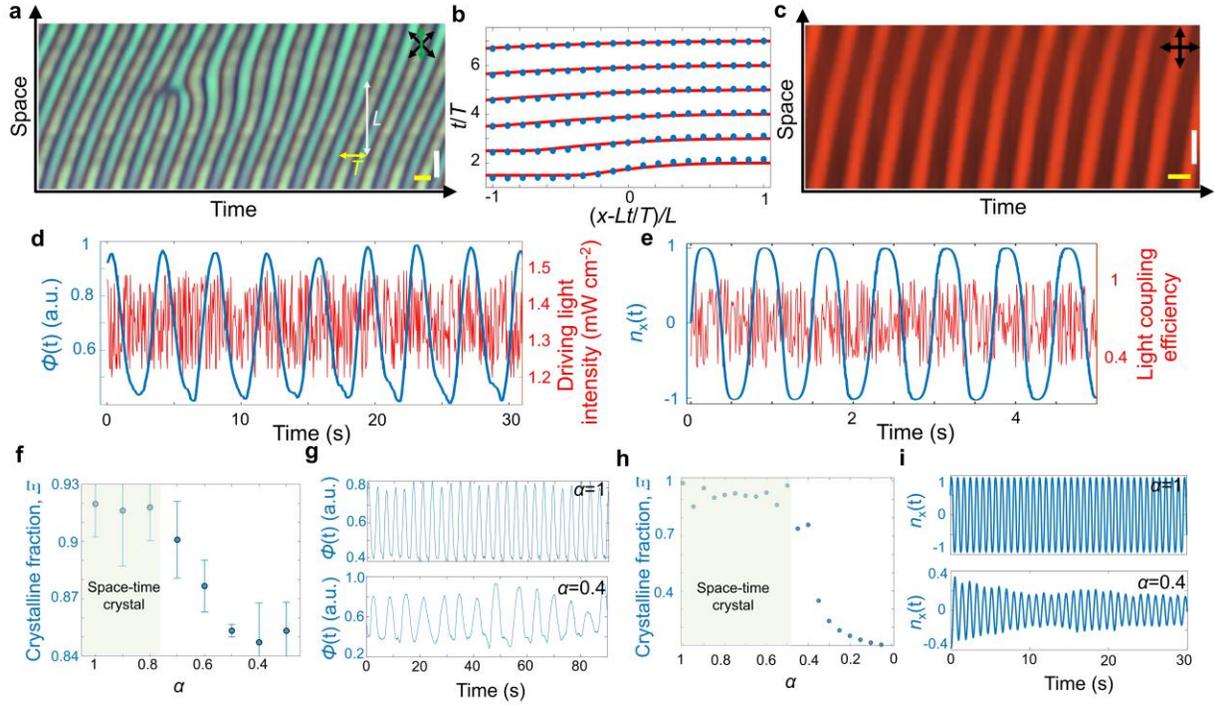

**Fig. 3 | Robustness of CSTCs against spatial and temporal perturbations. a**, Space-time image shows that the CSTC order recovers from an emergence of space-time dislocation, with the stripe pattern's slope of $\tan^{-1} L/T$. **b**, Layer displacement profile prediction (red solid lines) fits the experimental results (blue dots) reconstructed from (a). The horizontal coordinates are derived from the spatial coordinate $x$ minus $tL/T$. **c**, Space-time image of the CSTC under temporal perturbations. White scale bars are 10μm and yellow scale bar are 3s in (a,c). The retardation plate's slow axis is labelled by the green double arrow and crossed polarizers are labelled by black double arrows. **d**, Normalized light signals $\Phi(t)$ when temporally randomizing the driving light intensity. The time interval for each random step is 0.1s. a. u., arbitrary units. **e**, A simulated realization obtained after temporally randomizing the light coupling efficiency. The time interval for each random step is 0.01s. **f**, Experiment measured crystal fraction $\Xi=\sum_{1/T-\delta}^{1/T+\delta}\Phi(\omega/2\pi)/\sum\Phi(\omega/2\pi)$ versus $\alpha$ with error bars indicating standard deviations from 10 realizations, the driving light intensity is randomly distributed in $[\alpha,1]W_{\text{driving}}$, where $W_{\text{driving}}=1.5$ mWcm$^{-2}$ and $\delta=0.03$Hz. **g**,

Normalized light signals $\Phi(t)$ for $\alpha=1$ (top) and $\alpha=0.4$ (bottom). **h**, Computer simulated $\Xi=\sum_{1/T-\delta}^{1/T+\delta} n_x(\omega/2\pi) / \sum n_x(\omega/2\pi)$ versus $\alpha$, with the light coupling efficiency randomly distributed in $[\alpha,1]\eta_{max}$, where $\eta_{max}=0.5$ and $\delta=0.03$Hz. **i**, Simulations for $\alpha=1$ (top) and $\alpha=0.4$ (bottom).

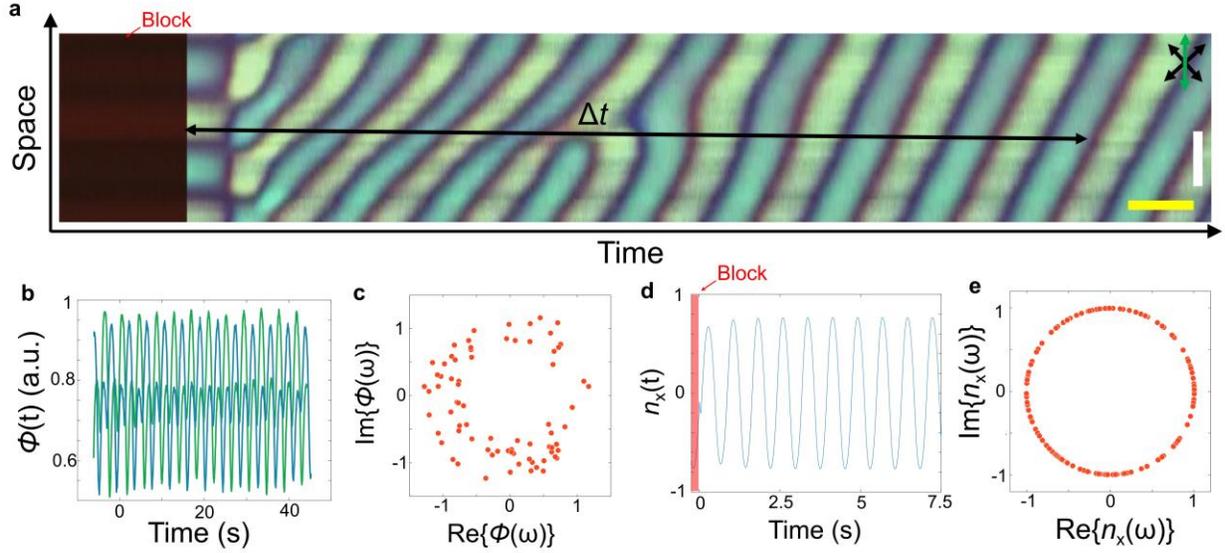

**Fig. 4 | Spontaneous symmetry breaking of CSTCs. a**, Space-time image of an experimental realization to measure the relative time phase. We first block the driving light with red colour filter, then calculate the relative time phase with the light signal sequence $\Phi(t)$ after a time interval $\Delta t$. White scale bar is 10μm and yellow scale bar is 3s. The retardation plate's slow axis is labelled by the green double arrow and crossed polarizers are shown by black double arrows. **b**, Normalized light signals captured within a small area versus time from two experimental realizations, which have a time phase difference about π. a. u., arbitrary units. **c**, Experimentally measured distribution of the relative time phases. **d**, A simulated realization that shows the evolution of *x* component of the director after blocking-unblocking the external drive. **e**, Simulated distribution of relative time phases calculated while using the *x* component of **n(r)**.

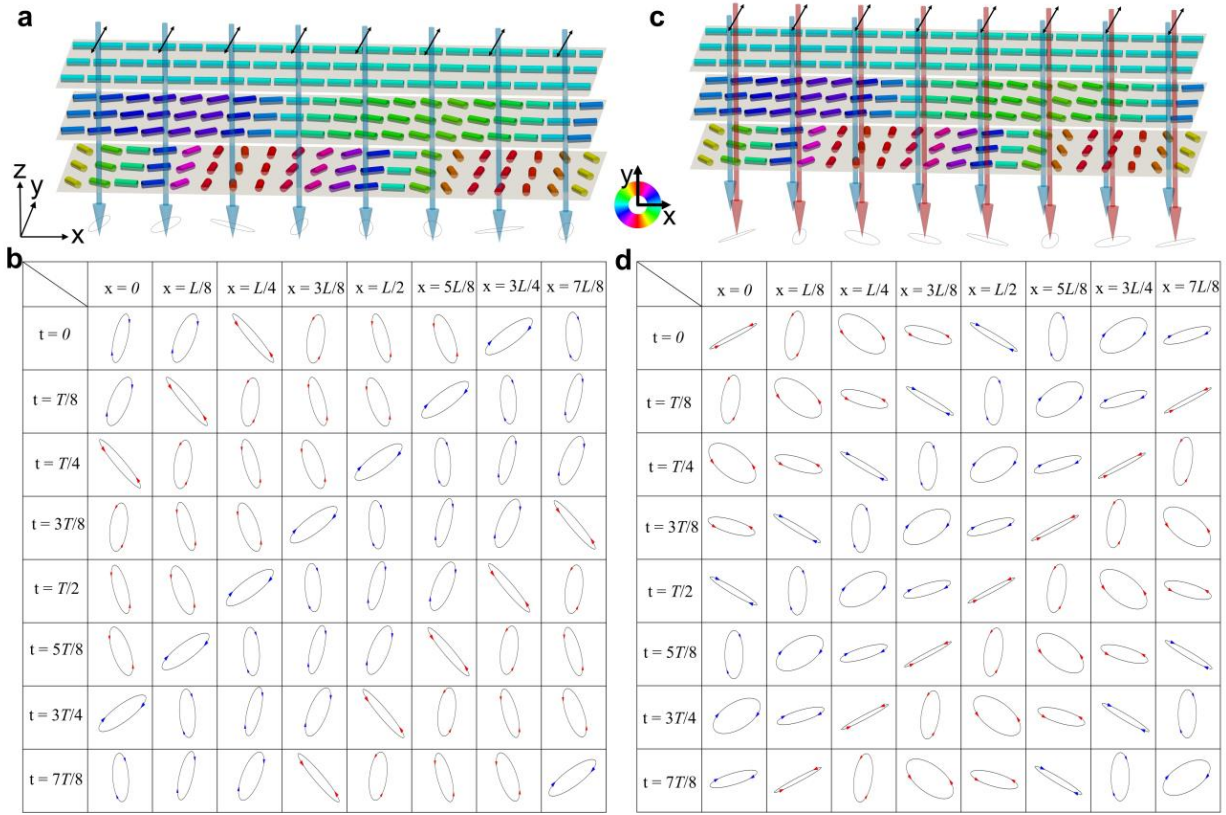

**Fig. 5 | Phase accumulation versus space-time coordinates. a**, Schematic of the phase accumulation process when the blue driving light (450nm, shown by blue arrows) passes through the cell. **b**, Spatial and temporal distribution of the output light polarizations within one temporal (*T*) and spatial (*L*) period, as the 450nm linearly-polarized input driving light passes through the cell. **c**, Schematic of the phase accumulation process when the modulated light (1300nm, shown by red arrows) passes through the cell. Linear and elliptical polarization states of the light in (a) and (c) are marked by the black double arrows and ellipses, respectively. Cylinder colours represent azimuthal angles of **n(r)** orientation defined by the scheme in the right inset of (a). **d**, Spatial and temporal distribution of the output light polarizations within one *T* and *L* periods, as the 1300nm linearly-polarized input modulated light passes through the cell. In (b) and (d), left- and right-handed elliptical polarization states are marked by blue and red arrows, respectively.

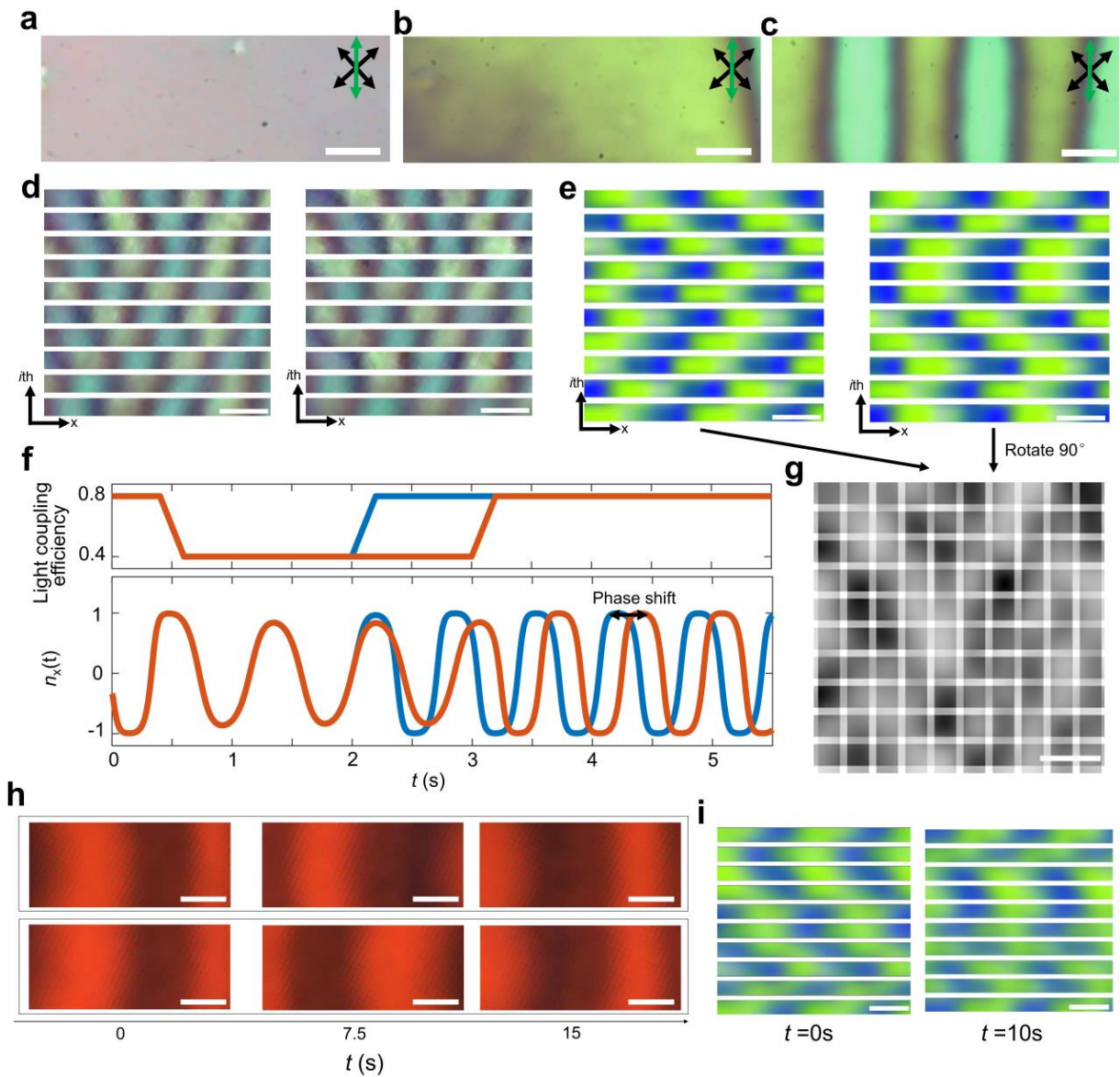

**Fig. 6 | CSTC applications in anti-counterfeiting designs. a-c**, Polarizing optical micrographs of disordered states (a,b) and the CSTC of spatiotemporal topological solitons (c) that can serve as a "time watermark". **d**, Two spontaneously formed experimental fingerprint states assembled from multiple CSTCs. **e**, Another pair of spontaneously formed numerically simulated fingerprint states assembled from multiple CSTCs. These fingerprint states are non-identical and the vertical axis in (d,e) represents individual CSTCs. **f**, By smoothly changing the light coupling efficiency (top) in different realizations (plot with blue and red, respectively), the CSTC phase, shown by the

evolution of x component of **n(r)**, can be tuned to any desired angle (bottom); the phase difference depends on the tunning process. **g**, A snapshot of a 2+1D barcode created by superimposing two fingerprint states, displayed by grayscale. Using the phase tuning method, information can be encoded and stored in this 2+1D barcode. **h**, Polarizing optical micrographs of two CSTCs, captured over time. The temporal periodicity of top row CSTC is 3.48s while that of the bottom row is 4.54s. Thus, after ~15s (14.90s), the phase of top CSTC aligns with that of the bottom one. **i**, Fingerprint states assembled from numerically simulated CSTCs over a 10s interval, with each row's CSTC having a different *T*. The right state cannot be generated by shifting the left fingerprint state by the same phase angle. However, if the temporal periodicity of CSTCs in each row is known, the right state can be predicted from the left fingerprint state. White scale bars are 10μm in (a-e,g-i).

**Extended Data Figures**

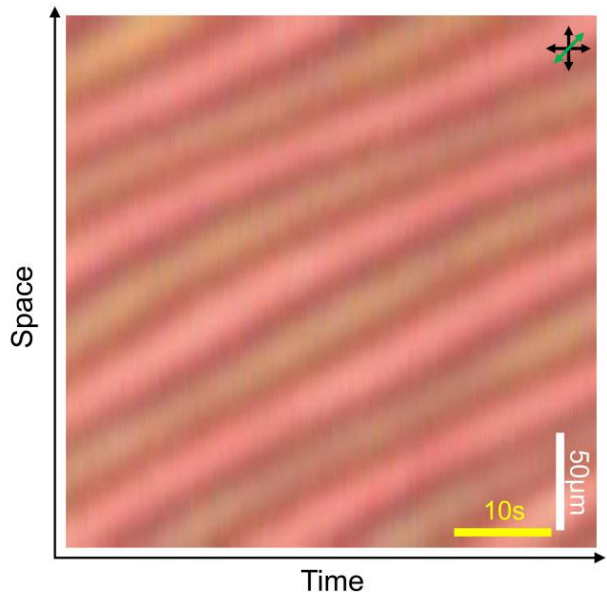

**Extended Data Fig. 1 | CSTC in a thick cell.** Space-time image of a CSTC obtained for the cell thickness $d = 4$ μm. White scale bar indicates 50μm; yellow scale bar indicates 10s. The retardation plate's slow axis is labelled by the green double arrow and crossed polarizers by black double arrows.

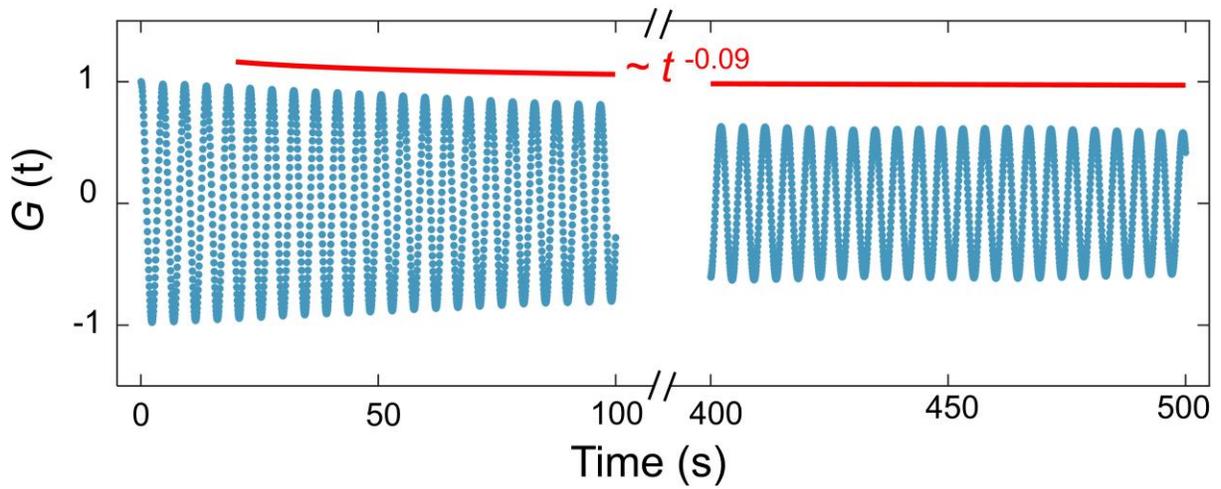

**Extended Data Fig. 2 | Quasi-long-range order of a CSTC.** Correlation function $G(t)$ versus time. The red solid line is the fit of power law decay, with an exponent of -0.09, indicating quasi-long-range order in time.

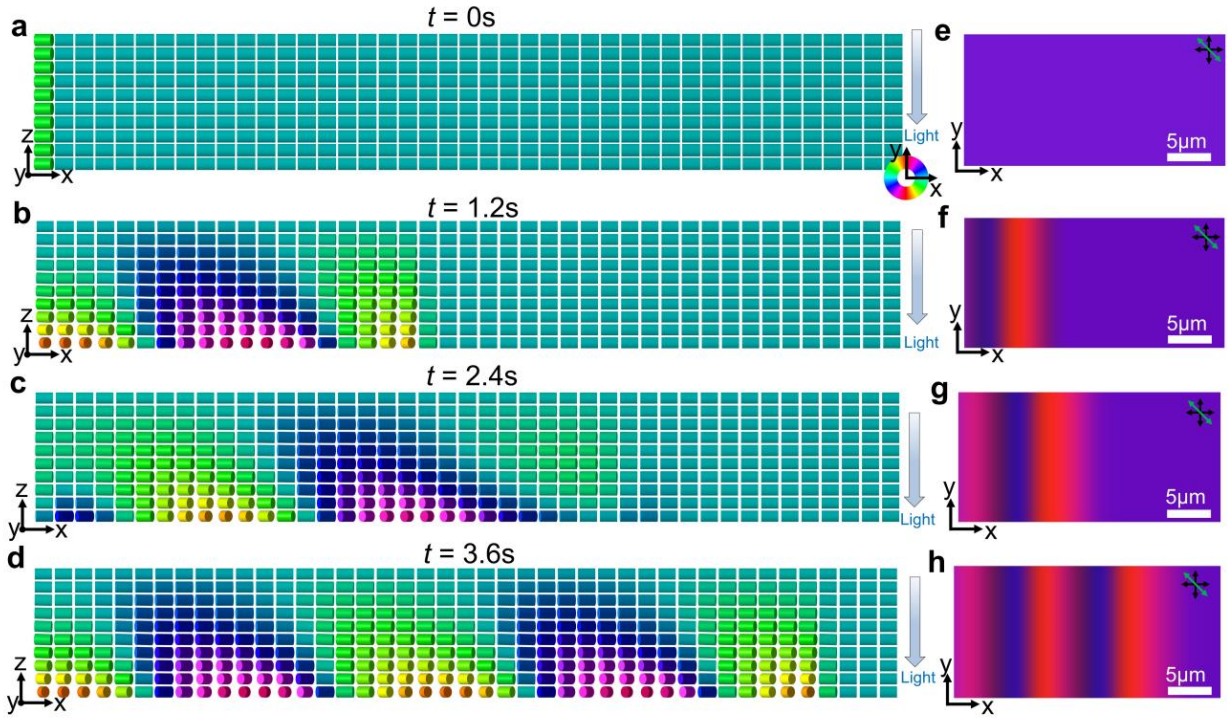

**Extended Data Fig. 3 | Spontaneous emergence of the CSTC in computer simulations. a**, Director field **n(r)** in the *x-z* cross-section that has a small perturbation on the left side. The structure is invariant along the *y* direction. **b-d**, Simulated director field **n(r)** shows the spontaneous emergence of spatially periodic structures stemming from a small perturbation starting from (a). The cylinder colours represent azimuthal angles of the director orientation as defined by the coloured circle in the bottom right of (a). The elapsed times are marked on their top and the light propagation directions are marked on their right, respectively. **e-h**, Simulated polarizing optical micrographs corresponding to the structures of director field of (a-d), respectively. The retardation plate's slow axes are labelled by green double arrows and crossed polarizers by black double arrows.

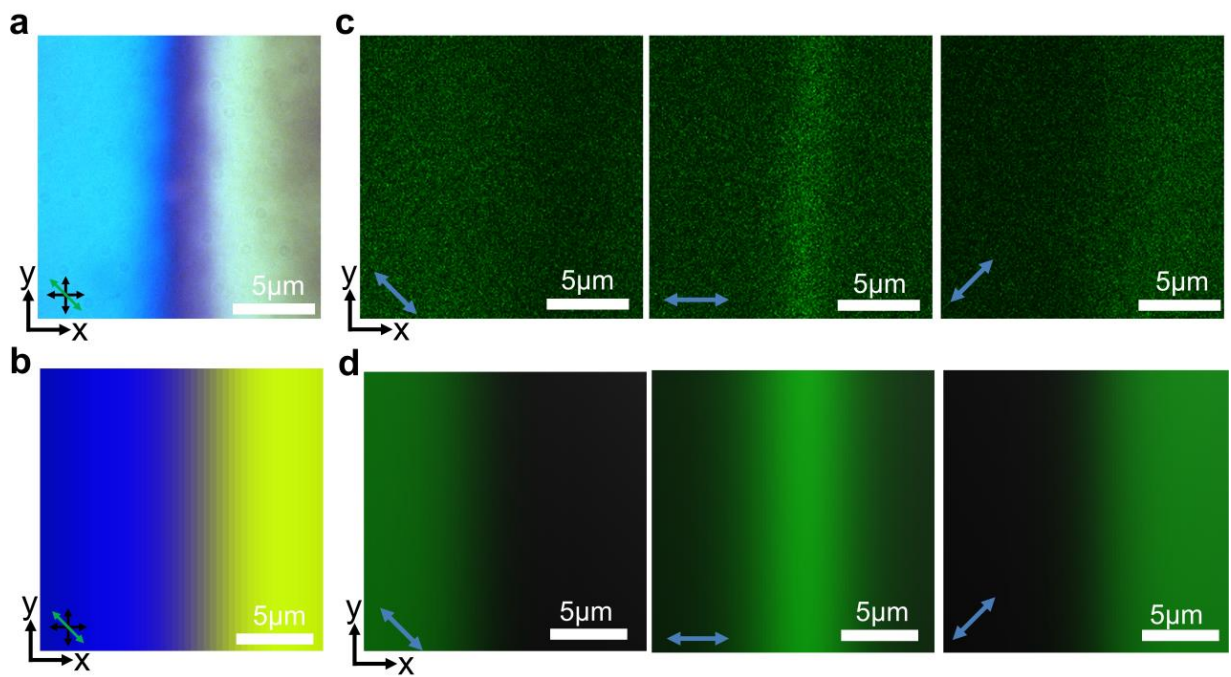

**Extended Data Fig. 4 | Polarizing optical micrographs and three-photon excitation fluorescence polarizing microscopy of CSTCs. a**,**b**, Experimental (a) and numerically simulated (b) polarizing optical micrographs of CSTCs. The slow axes of a phase retardation plate are labelled by green double arrows and crossed polarizers by black double arrows. **c,d**, Experimental (c) and numerically simulated (d) three-photon excitation fluorescence polarizing microscopy images obtained for different linearly polarized femtosecond laser excitations, with marked excitation light polarizations (blue double arrows). White scale bars indicate 5μm.

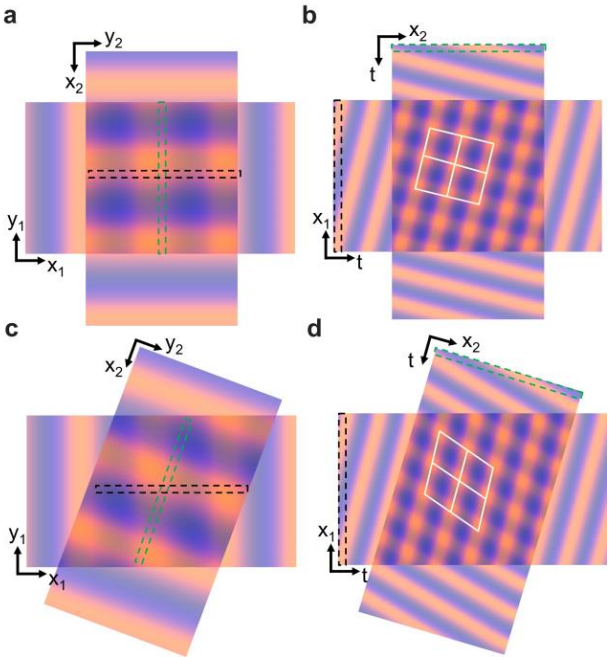

**Extended Data Fig. 5 | Time-crystalline diversity from superposition of two CSTCs. a,** An orthorhombic lattice pattern generated by orthogonally superimposing polarizing optical micrograph snapshots of two CSTCs. **b**, An orthorhombic lattice pattern derived from space-time images of two CSTCs, where the selected area is marked in (a) with dashed rectangles. **c,** A monoclinic lattice pattern generated by obliquely superimposing polarizing optical micrograph snapshots of two CSTCs. **d**, A monoclinic lattice pattern produced from space-time images of two CSTCs, where the selected area is marked in (c) with dashed rectangles. The 2×2 white lattices in (b) and (d) correspond to the orthorhombic and monoclinic lattices, respectively. Spatial and temporal axes are labeled in each image.

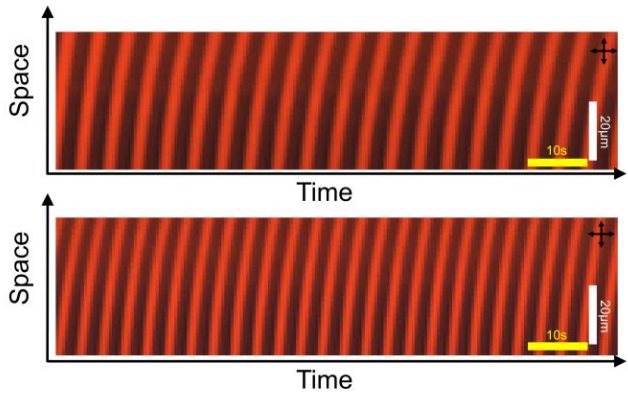

**Extended Data Fig. 6 | CSTCs driven by different driving light intensities.** Space-time images of CSTCs with different temporal periodicity driven by the low (top) and high (bottom) intensity, respectively. White scale bars indicate 20μm and yellow scale bars indicate 10s; the crossed polarizers are labelled by black double arrows.

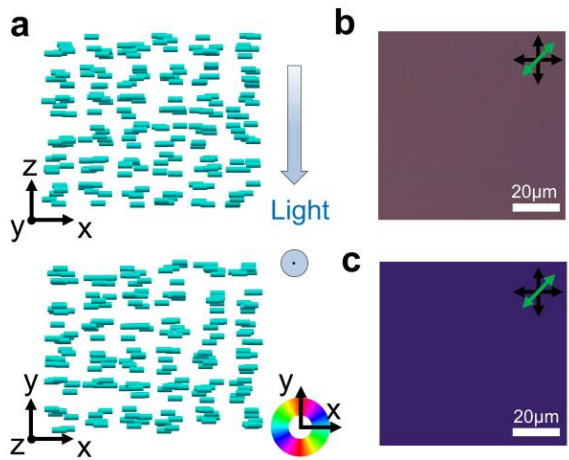

**Extended Data Fig. 7 | The time symmetry unbroken state. a**, Simulated director field of the TSU phase. The structure is invariant along the *y* direction, and the illustrated *x-y* cross-section of director field is at the mid-plane along *z* direction of the sample. The polarization of the linearly polarized light is along *y* direction at the top surface, and the propagation direction is marked on the right. The cylinders are coloured based on the azimuthal angles of the director, as defined by the coloured circle on the bottom right. **b,c**, Experimental (b) and numerically simulated (c) polarizing optical micrographs of the TSU phase. White scale bar indicates 20μm; the slow axes of a phase retardation plate are labelled by green double arrows and crossed polarizers by black double arrows.

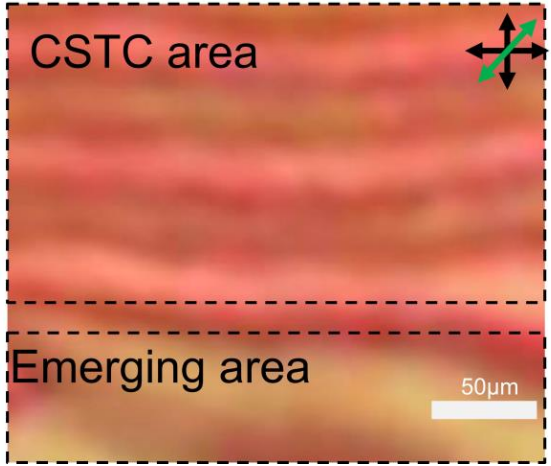

**Extended Data Fig. 8 | Polarizing optical micrograph shows the CSTC and emerging area.** The CSTC area shows a good spatial and temporal periodicity, which appears from the emerging area. White scale bar indicates 50μm. The slow axis of a phase retardation plate is labelled by the green double arrow; crossed polarizers by black double arrows.

**16**, 454–461 (2022).

# Methods

**Materials and sample preparation**

The glass substrates are coated with photo-responsive material 2-(4-dimethylamino-phenylazo)-N-(3-triethoxysilane-propyl)-benzamide (dMR)[35,56], which is sensitive to the blue and violet light and insensitive to the red light. To coat monolayer dMR on the glass surfaces, we submerge the glass plates into a 1 wt% solution of dMR in toluene at temperature of 45 °C. After a 90-min submersion, the dMR molecules are bonded to the glass surfaces, we then wash away the excess dMR by a toluene rinse, followed by blowing the glass plates with dry nitrogen and curing them at 115 °C for 2 hours. The LC cells are constructed using two glass substrates coated with monolayers of dMR, where the cell thickness $d$ = 2-4μm is defined by glass spheres mixed with a methanol-diluted epoxy. Once the epoxy has cured, we fill the cell via capillary forces with nematic 4-cyano-4'-pentylbiphenyl (5CB, EM Chemicals).

**Optical imaging and video microscopy**

Polarizing optical micrographs and videos are obtained with a multi-modal imaging setup built around an IX-81 Olympus inverted optical microscope (which is integrated with the three-photon excitation fluorescence polarizing microscopy imaging setup described below) and an IX-83 inverted optical microscope (which is integrated with the driving light intensity control system), with charge-coupled-device cameras (Grasshopper, Point Grey Research). High-numerical-aperture (NA) Olympus objectives 100×, 40×, 20× and 10× with NA =1.4, 0.75, 0.4 and 0.4, respectively, are used. The schematics of the used microscopes are shown in Fig. 1a and Supplementary Fig. 4. When recording the videos of CSTCs, we noted that within some small regions, the light signals may vary periodically over time while having no quasi-long-range temporal order (Supplementary Fig. 5) as compared to the CSTCs, which is expected due to the lack of many-body interactions.

Nonlinear optical imaging of the CSTC structures is performed using the three-photon excitation fluorescence polarizing microscopy setup built around the IX-81 Olympus inverted optical microscope[57]. We use a Ti-Sapphire oscillator (Chameleon Ultra II; Coherent) operating at 870 nm with 140-fs pulses at an 80 MHz repetition rate, as the source of the laser excitation light. An oil-immersion 100× objective (NA = 1.4) is used to collect the fluorescence signals, which are detected by a photomultiplier tube (H5784-20, Hamamatsu) after a 417/60-nm bandpass filter. With a third-order nonlinear process, the LC molecules are excited via the three-photon absorption process and the signal intensity scales $\propto \cos^6\beta_0$, where $\beta_0$ is the angle between the polarization of the excitation light and the long axis of the LC molecule. We utilize a half-wave plate to control different polarization states of the excitation. When **n(r)** is nearly parallel to the polarization of the laser beam, large $\cos\beta_0$ corresponds to the strong three-photon excitation fluorescence polarizing microscopy signal intensity. Computer simulations of the three-photon excitation fluorescence polarizing microscopy images are also based on the signal intensity. The 2D cross-sections, produced from both experimental and numerical data, for different linear polarizations of excitation light, are shown in Extended Data Fig. 4c,d.

**Quasi-long-range order and relative time phases of the CSTC**

To obtain the time order of CSTCs, we calculate the correlation function $G$ in time coordinate, which is a common tool for analyzing the spatial order of crystals and liquid crystals. For crystals, the spatial correlation function $G(r)$ is a constant, where $r$ is the distance between the two measured positions. For smectic liquid crystals, the spatial correlation function $G(r)$ decays as $\sim r^{-\zeta}$ ($\zeta<0.15$) along the smectic layers, which is a quasi-long-range order[37]. For CSTCs, we measure the normalized digital signal $\Phi_i(t)$ of each pixel at different times, where subscript $i$ denotes spatial coordinates. The correlation function $G(t)= \sum_i G_i(t)= \sum_i \langle\Phi_i(t)\Phi_i(0)\rangle-\langle\Phi_i(t)\rangle\langle\Phi_i(0)\rangle$ is calculated with 2200 spatial pixels and 9000 temporal frames, showing a quasi-long-range order in time. In experiment, the relative time phases are measured from 75 experimental realizations. In each realization, the driving light is blocked with red colour filter at first (allowing only red colour light pass through). Subsequently, we remove the red colour filter, and the CSTC spontaneously emerges. After a time interval $\Delta t$ ($\Delta t = 60$s), we start measuring the light signals from the recorded video, and calculate the phase using Fast Fourier Transform analysis function in MATLAB (MathWorks).

**Numerical modelling of solitonic quasi-particles and their crystals**

The elastic energy cost of spatial deformations in the bulk of $\mathbf{n}(\mathbf{r})$ is expressed in Eq. (1) and the surface energy is expressed in Eq. (2). The total Frank-Oseen free energy $F$ is the sum of these two terms, and a relaxation routine based on the variational method is used to identify an energy-minimizing configuration $\mathbf{n}(\mathbf{r})$. At each iteration, the $\mathbf{n}(\mathbf{r})$ is updated based on a formula derived from the Euler-Lagrange equation:

$$\mathbf{n}_i^{\text{new}} = \mathbf{n}_i^{\text{old}} - \frac{\text{MSTS}}{2}\frac{\delta F}{\delta \mathbf{n}_i}, \quad (3)$$

where subscript $i$ denotes spatial coordinates, $\frac{\delta F}{\delta \mathbf{n}_i}$ denotes the functional derivative of the total free energy with respect to $\mathbf{n}_i$, and MSTS is the maximum stable time step of the minimization routine, determined by the elastic constants and the spacing of the computational grid. To scale the time step in the real system, we assume that the director is governed by the balance equation: $\frac{\delta F}{\delta \mathbf{n}_i} = -\gamma \frac{\partial \mathbf{n}_i}{\partial t}$, and combining with equation (3) we have the time interval $\Delta \tau = \frac{\text{MSTS}}{2}\gamma$ for each iteration. After obtaining $\mathbf{n}(\mathbf{r})$ at each iteration, we calculate the electric and magnetic fields of the driving light (450 nm) traversing through the LC, which is linearly polarized initially. At the top surface (Fig. 2a), the light does not interact with the LC field, so the polarization remains along the initial direction, $y$ direction, and drives the surface director $\mathbf{n}_s$ to orient to the $x$ direction. When it passes through the cell and interacts with the LC, because of the birefringence of the LC host medium (Supplementary Table 2), the linearly polarized light becomes elliptically polarized light. After it passes through the LC's bulk, the surface director at the bottom surface (Fig. 2a) is driven to orient perpendicular to the direction of electric field (real part) of Jones vector. With the new orientation of $\mathbf{n}_s$, $\mathbf{n}(\mathbf{r})$ is updated in the next iteration.

The initial condition of $\mathbf{n}(\mathbf{r})$ comes from the spontaneous emergence of the CSTC (Extended Data Fig. 3). To avoid the edge effects on the lateral sides, we apply the periodic boundary condition (with three repeat units along $x$ direction, Fig. 2g) to obtain the continuous space-time crystallization phase. The 3D spatial discretization is performed on 3D square-periodic $120 \times 10 \times 10$ grids, and the spatial derivatives are calculated using finite-difference methods with the second-order accuracy, we find that double or triple the system size does not affect the temporal periodicity. In simulations, we use the physical parameters of 5CB as the model LC also used in experiments. The temperature dependent parameters $K_{11}$, $K_{22}$, $K_{33}$, and $\Delta n$ (birefringence) are fitted with the empirical equations:

$$\Omega = A_e + B_e(\Delta T_{em} - T_e)^{1/2} + C_e(\Delta T_{em} - T_e) + D_e(\Delta T_{em} - T_e)^{3/2}, \quad (4)$$

where $\Delta T_{em}$ denotes the difference between the nematic–isotropic transition temperature $T_{NI}$ = 35.1 °C and the LC temperature, $A_e$, $B_e$, $C_e$, $D_e$ and $T_e$ are fitting coefficients and $\Omega$ correspond to the temperature dependent parameters (Supplementary Tables 1,2 and Supplementary Fig. 1). For all simulations, $d$ = 3 μm, $\gamma$ = 77 mPas and $W$ = $10^{-5}$ J m$^{-2}$.

The polarizing optical micrographs are simulated by means of the Jones-matrix method, using the configurations of **n(r)** at a certain moment for the CSTCs. Briefly, we first split the cell into 10 thin sublayers along the $z$ direction, then calculate the Jones matrix for each pixel in each sublayer by identifying the local optical axis and ordinary and extraordinary modes' phase retardation, originating from optical anisotropy. We obtain the Jones matrix for the whole LC cell by multiplying all Jones matrices corresponding to each sublayer, and a 530nm retardation plate is included for an extra Jones matrix. The simulated single-wavelength polarizing optical micrograph is obtained as the respective component of the product of the Jones matrix and the incident light's polarization. To properly reproduce the polarizing optical micrographs observed in experiments, we produced images separately for three different wavelengths spanning the visible spectrum (450, 550, and 650 nm) and then superimposed them, according to experimental light source intensities at the corresponding wavelengths.

**Methods-only references:**

**Additional Information:** Supplementary Information is available for this paper. This Supplementary Information file includes Supplementary Figures 1-5, Supplementary Tables 1,2 and Supplementary Videos 1-7.